\documentclass[aps,prl,twocolumn,preprintnumbers,superscriptaddress,nofootinbib]{revtex4-1}
\usepackage{natbib}
\usepackage[utf8]{inputenc}
\usepackage{lmodern}
\usepackage{amsmath,amssymb}
\usepackage{graphicx}
\usepackage{url}
\usepackage{color} 
\usepackage{enumitem}
\usepackage{subfigure}
\usepackage[dvipsnames]{xcolor}
\usepackage[colorlinks=true,breaklinks=true]{hyperref}
\hypersetup{allcolors=[rgb]{0.0 0.0 0.6},linkcolor=[rgb]{0.75 0.05 0.05}}
\usepackage{bm}
\usepackage{amsmath}
\usepackage{amssymb}
\usepackage{slashed}
\usepackage{color}
\usepackage{accents}

\newcommand{\exclude}[1]{}

\newcommand{\be}{\begin{equation}}
\newcommand{\ee}{\end{equation}}
\newcommand{\beq}{\begin{equation}}
\newcommand{\eeq}{\end{equation}}
\newcommand{\bea}{\begin{eqnarray}}
\newcommand{\eea}{\end{eqnarray}}

\newcommand{\TT}{{\cal T}}

\newcommand{\PP}{{\cal P}}

\newcommand{\RR}{{\cal R}}

\newcommand{\ga}{\gamma}

\newcommand{\La}{\Lambda}
\newcommand{\si}{\sigma}

\newcommand{\cd}{\cdot}

\newcommand{\fNL}{f_\mathrm{NL}}

\definecolor{dgreen}{rgb}{0,0.6,0.0}

\begin{document}
\preprint{} 

\title{Constraining the bispectrum from bouncing cosmologies with Planck}
\author{Bartjan van Tent} 
\affiliation{Universit\'e Paris-Saclay, CNRS/IN2P3, IJCLab, 91405 Orsay, France.}
\author{Paola C. M. Delgado} 
\affiliation{Faculty of Physics, Astronomy and Applied Computer Science, Jagiellonian University,
30-348 Krakow, Poland.} 
\author{Ruth Durrer}
\affiliation{Department of Theoretical Physics, Universit\'e de Gen\`eve, Quai
E. Ansermet 24, Gen\`eve, 1211, Switzerland.}

\begin{abstract}
Bouncing models of cosmology, as they arise e.g.\ in loop quantum cosmology, can be followed by an inflationary phase and generate close-to-scale-invariant fluctuation spectra as observed in the Cosmic Microwave Background (CMB). However, they are typically not Gaussian and also generate a bispectrum.  These models can help to mitigate the large-scale anomalies of the CMB by considering substantial non-Gaussianities on very large scales, which decay exponentially on sub-horizon scales. It was therefore thought that this non-Gaussianity would not be visible in observations, which can only probe sub-horizon scales. In this letter we show that bouncing models with parameters such that they can significantly mitigate the large-scale anomalies of the CMB are excluded by the Planck data with high significance of, depending on the specific model, 5.4, 6.4 or 14 standard deviations.

\end{abstract}
\maketitle


{\it Introduction ~~~} The most commonly accepted idea for the generation of initial fluctuations in cosmology is inflation, which was pioneered in \cite{inflation1,inflation4,Mukhanov:1982nu}. Initially, inflation was invoked to solve the cosmological horizon and flatness problems~\cite{inflation2,inflation3}. However, since inflation cannot solve the singularity problem and since the flatness and horizon solutions are `post-dictions' of inflation, it is usually considered that the nearly scale-invariant and nearly Gaussian initial fluctuations are the most significant signatures of inflation. Furthermore, many simple inflationary models also predict a similar amount of tensor fluctuations with a tensor-to-scalar ratio of $r\gtrsim 0.1$. Present data~\cite{Tristram:2021tvh}, however, constrain this ratio to $r<0.032$, excluding many inflationary models.  Even though there are inflationary models compatible with present data, e.g., Starobinsky inflation~\cite{inflation1} or Higgs inflation~\cite{Bezrukov:2007ep}, it is always important to study whether alternatives to inflation can also lead to predictions which are compatible with observations.

One possibility to solve the singularity problem are `bouncing models', where the observed expanding Universe emerges from a collapsing phase. These models have a long history starting with Tolman~\cite{cbounce1}. Not always, but in many cases, they require a violation of the dominant energy condition to allow an increase of the Hubble parameter. Bounces can also be nonsingular, see e.g.~\cite{cbounce2,Easson:2011zy,cbounce6,cbounce8,Dobre:2017pnt}.
Especially attractive singularity-free bouncing cosmologies arise in loop quantum cosmology (LQC)~\cite{qbounce5,qbounce10}. A comparison of inflationary and bouncing cosmologies with respect to their performance in view of the Planck data is given in~\cite{ob0,ob2}.  Bouncing models in general do predict larger non-Gaussianities than inflationary models. The non-Gaussianity generated in  LQC has been investigated in~\cite{PhysRevD.97.066021}.

One of the most debated problems of standard cosmology are the large-scale anomalies of the CMB data,  
most importantly the power suppression on large scales and the dipolar asymmetry seen in the preference for odd-parity correlations~\cite{Kim:2010st,anomalies,Planck-nongaussian}. Even though these anomalies have a statistical significance around 2 to 3$\si$ and may be accepted as coincidences, they would be less `anomalous' in a model with significant non-Gaussianity on very large scales. And this is exactly what  LQC and the bouncing models investigated in \cite{Agullo:2020cvg} predict. In the present letter, these bouncing models, which are followed by a phase of slow-roll inflation, are studied and the amplitude $\fNL$ of the non-Gaussianity in the model is chosen such that the large-scale power suppression in the CMB has a p-value of about 20\%. 

In \cite{Agullo:2020cvg} it is argued that the exponential decrease of the non-Gaussianity on sub-horizon scales is sufficient to make it invisible in e.g.\ the CMB bispectrum, which gains most of its signal-to-noise from high $\ell$-values, which are well inside the horizon.
In \cite{Delgado:2021mxu} some of us have shown, using simple approximations, that the signal-to-noise ratio, $S/N$, of the requested non-Gaussianity is nevertheless substantial, and the signal should be visible in Planck.

In this letter we now investigate these models with the real Planck data using the binned bispectrum estimator derived in~\cite{Bucher:2009nm,Bucher:2015ura}. We determine the central value and the error bars of $\fNL$ for the bispectrum shapes proposed in~\cite{Agullo:2020cvg} from the data and find that there is no detection. Moreover, the values of $\fNL$ required in order to remove the anomalies are excluded by 5.4$\si$, 6.4$\si$ and 14$\si$ for the three models considered. 


{\it The bispectrum~~~} The regular bouncing model described in~\cite{Agullo:2020cvg} generates the following dimensionless power spectrum, ${\cal P}_{\cal R}(k)$,  of curvature fluctuations in Fourier space:
\bea
{\cal P}_{\cal R}(k) &=& A_s\left\{ \begin{array}{cc} (k/k_i)^2(k_i/k_b)^q & \mbox{if }~k\leq k_i \\
 (k/k_b)^q & \mbox{if }~k_i< k\leq k_b\qquad \\
 (k/k_b)^{n_s-1}  &\mbox{if }~k> k_b \,.\end{array}\right. 
 \eea
 The  bispectrum\footnote{\label{f1}Note that the definition of $\fNL$ in \cite{Agullo:2020cvg,Delgado:2021mxu} differs by a factor $-2$ from the one used here. This explains for example why there is a factor $+3/5$ instead of $-6/5$ in the expression of the bispectrum in \cite{Delgado:2021mxu} and why our numbers in Table~\ref{table:1} differ by a factor of $-2$ from the ones in those papers. Here we follow the definition of $\fNL$ used in the Planck analysis and given for example in~\cite{Komatsu:2001rj}.} , $B(k_1,k_2,k_3)$, is
 \bea
 B(k_1,k_2,k_3) &=& -\frac{6}{5}(2\pi^2)^2\fNL\Bigg[\frac{{\cal P}_{\cal R}(k_1)}{k_1^3}\frac{{\cal P}_{\cal R}(k_2)}{k_2^3} + \qquad \nonumber\\
 &&  \hspace{-1cm}\frac{{\cal P}_{\cal R}(k_1)}{k_1^3}\frac{{\cal P}_{\cal R}(k_3)}{k_3^3}+\frac{{\cal P}_{\cal R}(k_3)}{k_3^3}\frac{{\cal P}_{\cal R}(k_2)}{k_2^3}\Bigg] \times
 \nonumber \\  &&  \qquad \exp\left(-\ga\frac{k_1+k_2+k_3}{k_b}\right)\,. \label{e:Bk}
\eea
Here $n_s=0.9659$ and $A_s=2.3424\times 10^{-9}$ are the amplitude and spectral tilt of the curvature perturbations measured by Planck~\cite{Planck:2018vyg}.  The scale $k_i$, set to $k_i=10^{-6}$~Mpc$^{-1}$, is a very large scale, below which perturbations are significantly suppressed. Our results are not sensitive to this scale. The scale $k_b=0.002$~Mpc$^{-1}$ is the pivot scale above which the bispectrum is exponentially suppressed.  Its value is related to $\fNL$. Making it smaller in order to suppress also lower $k$-values, we have to increase $\fNL$ to achieve the goal of removing the CMB anomalies. On the other hand, by making it larger we would obtain a power spectrum which no longer agrees with the Planck observations. We therefore choose the largest possible value for $k_b$ which is of the order of the smallest values of $k$ which are well measured in the CMB power spectrum observed by Planck. The parameters $q$, $\ga$ and $\fNL$
depend on the  bounce, see \cite{Agullo:2020cvg,Delgado:2021mxu} for details. Their values for the models studied in this work are shown in Table~\ref{table:1}. The parameters of model~2, with $q=-0.7$, correspond to LQC while model~3, with $q=-1.24$, is a phenomenological bouncing model which provides
the best fit to the Planck data in a Markov Chain Monte Carlo (MCMC) analysis performed with Planck TT and low-$\ell$ EE power spectra carried out in \cite{Agullo:2020cvg}.
The fit is excellent, even somewhat better than $\La$CDM. This value is also close to the smallest value of $q$ which can still resolve the large-scale anomalies as we require here. The value of $|\fNL|$ needed in this model is significantly smaller. Finally, we also study a somewhat larger value than the one of LQC, $q=-0.5$, which correspondingly requires a larger value of $\fNL$ to resolve the large-scale anomalies.  We call this model~1. In all three cases we assume the smallest possible values for $\fNL$ such that the large-scale anomalies appear with a probability of 20\%.  This requires that the curvature scale of the bounce is the Planck scale. We also give the values of  $\fNL$ for the 10\% and 5\% probabilities. Note, however, that in standard $\La$CDM this probability is about 2\%, hence not so much smaller than the last value. The analysis in the next section is performed for $\fNL$ of 20\% in Table~\ref{table:1}; the results for the other probabilities can be obtained by linear rescaling.
\begin{table}
\begin{center}
\begin{tabular}{|c |c|c|c|c|c| } 
 \hline
   model&$q$ & $\gamma$ & $\fNL$ 20\%&   $\fNL$ 10\%& $\fNL$ 5\%  \\ 
 \hline \hline
 1& $-0.5$& 0.588 &  $-2516$ &$-1661$ & $-1283$\\
 \hline
 2& $-0.7$ & 0.6468 &  $-1663$ & $-1098$ & $-848$\\
 \hline
 3& $-1.24$ &  0.751 &  $-480$ & $-317$ & $-245$ \\ 
 \hline
\end{tabular}
\caption{The values of the parameters considered in this work. The $\fNL$ parameters are chosen according to \cite{Agullo:2020cvg} in order to alleviate the power suppression anomaly (but note the factor $-2$ difference in definition here as compared to~\cite{Agullo:2020cvg}, see footnote~\ref{f1}). We also give the values of $\fNL$ needed to obtain a probability of 10\% and 5\%, respectively, to observe the power suppression anomaly using the definition of~\cite{Agullo:2020cvg}.
}
\label{table:1}
\end{center}
\end{table}

The reduced CMB bispectrum is obtained in terms of the Fourier space bispectrum via~\cite{Durrer:2020fza}
\bea
B_{\ell_1\ell_2\ell_3} &=& \left(\frac{2}{\pi}\right)^3\int _0^\infty \!\! \!\! dx\, x^2 \!\int _0^\infty\!\! \!\! dk_1\!\int _0^\infty\!\! \!\! dk_2\!\int _0^\infty\!\! \!\! dk_3 \times \nonumber \\
&&\hspace{-1.3cm} \left[\prod_{j=1}^3\TT(k_j,\ell_j)j_{\ell_j}(k_jx)\right] (k_1k_2k_3)^2B(k_1,k_2,k_3)   \,, \qquad \label{eNG:Bzeta-blll}
\eea
where $\TT(k,\ell)$ is the CMB transfer function and $j_\ell$ is the spherical Bessel function of index $\ell$. 
In this expression the forward Fourier transform has no factors of $2\pi$ and the transfer function is defined such that the  CMB temperature power spectrum is given by
\be
C_{\ell} = 4\pi\int dk \, k^2(\TT(k,\ell))^2\PP_\RR(k) \,,
\ee
where $\PP_\RR$ is the dimensionless curvature power spectrum, see \cite{Durrer:2020fza} for more details. Note that the normalisation of the transfer functions depends on the definition. This transfer function, e.g., differs by a factor $\sqrt{\ell(\ell+1)/2}$ from the one given  in \cite{Hu:2003vp}.
\vspace{0.2cm}

{\it Limits from Planck~~~}
In a previous paper \cite{Delgado:2021mxu} some of us have estimated the CMB bispectrum induced by these bouncing cosmology models via rather crude analytic approximations. There we found that the models should have a signal-to-noise ratio in the Planck data of 25 to 50 and therefore be well detectable. In this work we compute the CMB bi\-spec\-trum exactly using the numerical transfer functions as determined by CAMB\footnote{\url{http://camb.info}} and
search for the signal in the truly observed Planck data. We employ the binned bispectrum estimator described in \cite{Bucher:2009nm,Bucher:2015ura} and used in the Planck analyses~\cite{Planck:2013wtn,Planck:2016aaa,Planck-nongaussian}. We analyze the cleaned CMB temperature and E-polarization maps of the Planck 2018 release, created by the SMICA component separation method~\cite{Planck:2018yye}, which have an angular resolution of 5'. We mask them using the common masks of the Planck 2018 analysis, which leave a sky fraction of 78\%. Error bars and linear correction terms are computed using 300 simulations. For more details about the data, see~\cite{Planck-nongaussian}.

\begin{figure}
\begin{center}
\includegraphics[width=8.5cm]{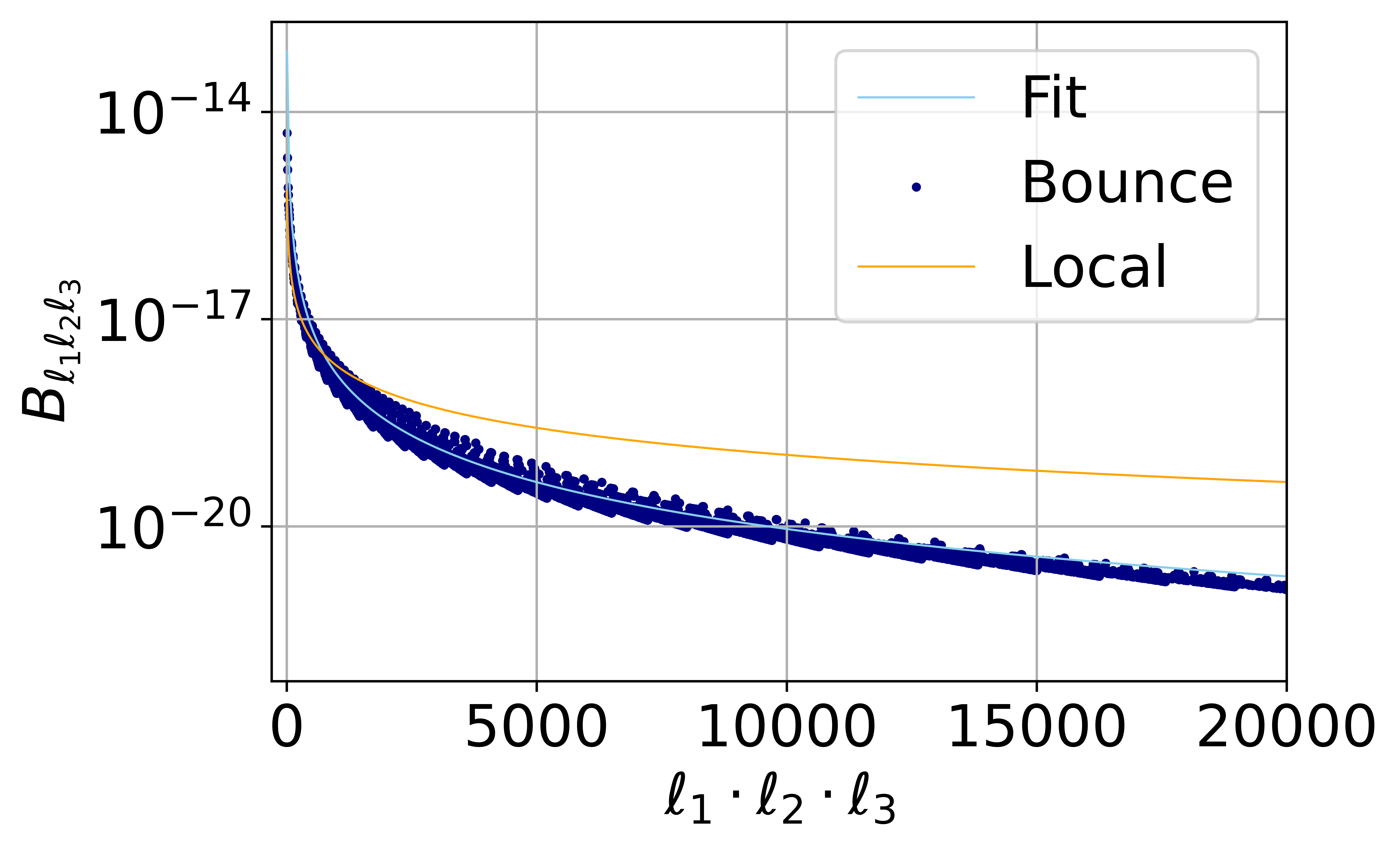}
\includegraphics[width=8.5cm]{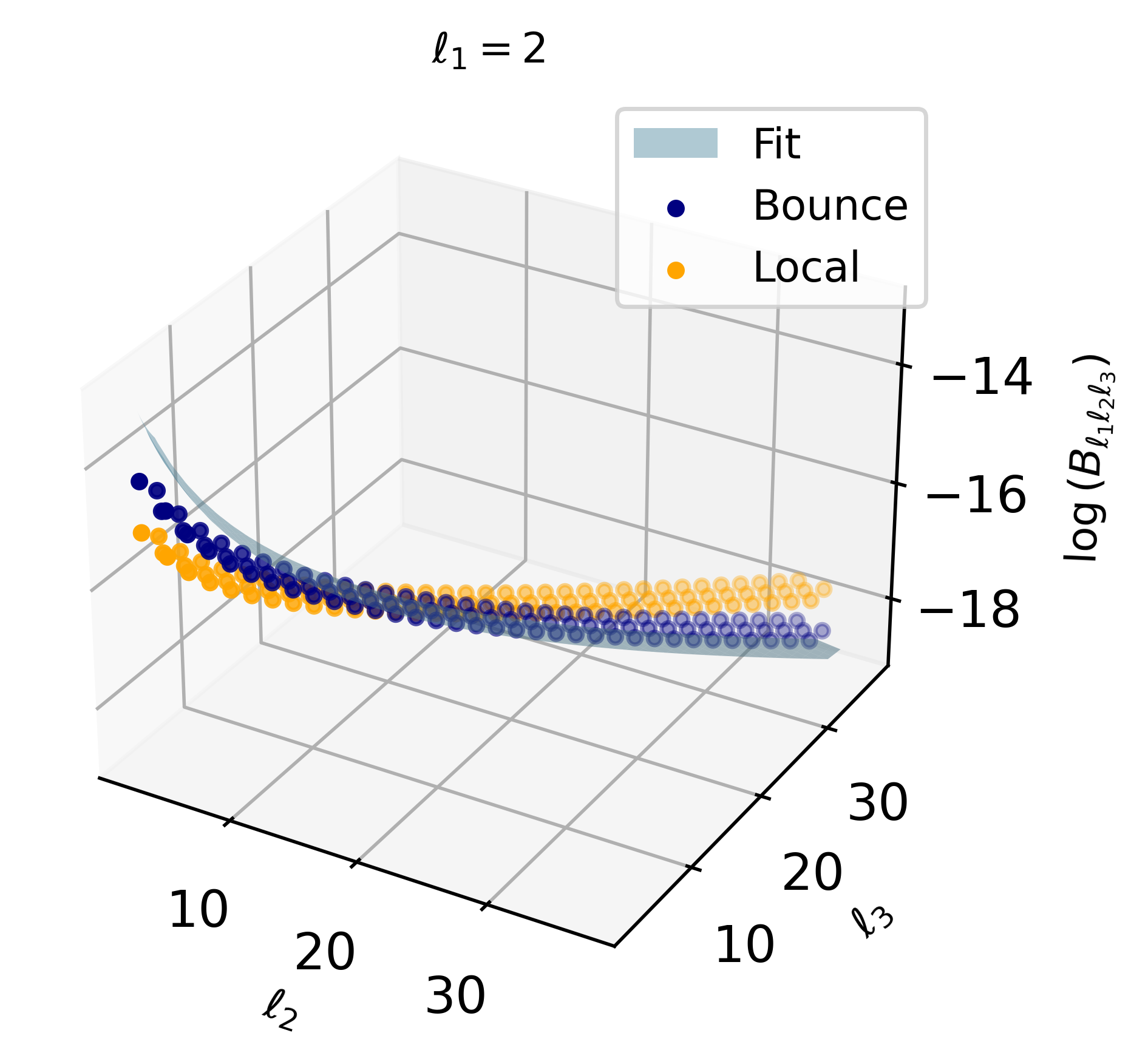}
\end{center}
\caption{\label{bg1} {\it Top panel:} The bouncing bispectrum computed with the numerical transfer functions (blue dots), fit to the bispectrum obtained in \cite{Delgado:2021mxu} (cyan) and the local bispectrum (yellow), for $q=-0.7$  (multiplication by $\fNL$ included). The bispectrum is plotted as a function of the product $L\equiv \ell_1 \ell_2 \ell_3$, which allows plotting all values of the 3D bispectrum in a 2D plot, at the price of having multiple $(\ell_1,\ell_2,\ell_3)$ configurations corresponding to the same value of the product $L$.\\
{\it Bottom panel :} The same bispectrum for $\ell_1=2$ fixed as a function of $\ell_2$ and $\ell_3$, compared to the local bispectrum with the same value for $\fNL$. Only values of $\ell_i$ which satisfy the triangle inequality are plotted. The fitting formula is indicated as a cyan surface. }
\end{figure}

 Figure~\ref{bg1} shows the comparison between the bispectrum fit from \cite{Delgado:2021mxu} and the exact numerical bispectrum computed in this paper. While there are obvious differences, we see that the fit gives a reasonable approximation, despite the shortcomings of the analytic approximations on which it was based. These shortcomings are for example the fact that the integrated Sachs-Wolfe effect was ignored, even though it is important at the lowest values of $\ell$ where this template peaks. Also the contributions from the acoustic peaks are not accounted for in~\cite{Delgado:2021mxu}. However, we expect these to be negligible due to the exponential decay of the bispectrum. Furthermore, the integration routine used in this first paper was different and computationally much more demanding so that it cannot be used efficiently with the full numerical transfer functions. In~\cite{Delgado:2021mxu}  simple fits for the bispectra as functions of the product $L\equiv \ell_1\cd\ell_2\cd\ell_3$ were introduced. While these capture well the overall shape of the numerical results, they somewhat overestimate it at high $L$ and also, more importantly, at the dominant lowest values of $L$.
 Here the analytical fit is just shown for illustration but it is not used in our analysis.

The bispectrum amplitude $\fNL$ is determined from the data by template fitting. The theoretical bispectrum template (\ref{eNG:Bzeta-blll}) determined from (\ref{e:Bk}) is multiplied by the observed bispectrum $B_{\ell_1 \ell_2 \ell_3}$ of the CMB and divided by the expected bispectrum variance (which in the case of weak non-Gaussianity is just a product of the three measured power spectra $C_{\ell_1} C_{\ell_2} C_{\ell_3}$), summing over all values of $\ell_1, \ell_2, \ell_3$. This expression must finally be multiplied by a  factor to normalize the inverse-variance weights, and this factor is exactly the expected variance of $\fNL$. In the case that polarization data is included as well as temperature data, the division by the variance becomes a multiplication with the inverse covariance matrix, and the sum is also over polarization indices. The whole expression for $\fNL$ can simply be viewed as the normalized inner product of the bispectrum template with the observed bispectrum of the CMB:
\be
{f}_\mathrm{NL}=\frac{\left\langle B^\mathrm{th},B^\mathrm{obs}\right\rangle}
{\left\langle B^\mathrm{th}, B^\mathrm{th}\right\rangle}.
\ee
In the simple case of temperature only and no binning, this inner product is given by
\be
\langle B^A, B^B \rangle = \sum_{\ell_1 \leq \ell_2 \leq \ell_3}
\frac{B^A_{\ell_1 \ell_2 \ell_3} B^B_{\ell_1 \ell_2 \ell_3}}{V_{\ell_1 \ell_2 \ell_3}},
\ee
where $V$ is the variance of the observed bispectrum, which depends on the noise and beam characteristics of the experiment. For the explicit definitions of the inner product in the case of binning or when polarization is included, as well as for other expressions and more detailed explanations, see e.g.~\cite{Bucher:2015ura}.

Computing the observed bispectrum for all values of $\ell_1, \ell_2, \ell_3$ is computationally too expensive, hence estimators must use approximations. The binned bispectrum estimator used in this paper makes the approximation that the bispectrum templates we are looking for are sufficiently smooth and slowly changing, that it is enough to only compute the average value of the bispectrum in each bin of $\ell$ values. This is a good approximation for the bouncing bispectrum under consideration: it was explicitly tested that the standard binning with 57 bins used for the Planck 2018 analysis \cite{Planck-nongaussian} gives a negligible increase in variance compared to the exact non-binned template. The Planck binning was determined by minimizing the increase in the theoretical variance for the local, equilateral and orthogonal shapes due to the binning, taking into account the noise and beam characteristics of the Planck experiment for both temperature and polarization.

The bouncing bispectrum template has the property that it decreases extremely fast as a function of $\ell$ because of the exponential factor in~(\ref{e:Bk}). It was shown that cutting off the analysis at $\ell_\mathrm{max}=36$ does not change the expected Fisher error bar at all compared to the $\ell_\mathrm{max}=2500$ used in the Planck analysis. However, in the actual data analysis it is still important to use this much higher $\ell_\mathrm{max}$ in order to disentangle the bouncing bispectrum from other sources of non-Gaussianity that are present in the data, like extra-galactic point sources and the lensing bispectrum. Table~\ref{overlap} gives the correlation coefficients of the bouncing template (for the three different values of $q$) with the standard primordial and foreground templates of the Planck analysis (for temperature only, in order to also show extragalactic point sources and the Cosmic Infrared Background). These correlations coefficients are defined as
\be
C_{IJ} = \frac{F_{IJ}}{\sqrt{F_{II} F_{JJ}}},
\ee
where $I$ and $J$ are indices labeling the templates and $F$ is the Fisher matrix defined as $F_{IJ} = \langle B^I, B^J \rangle$.
We see that, once the full Planck range is used, the correlation with all the other templates is very small (although the 3-5\% correlation with the orthogonal shape is not completely negligible). Not surprisingly, the correlation between the three bouncing templates, on the other hand, is very large. There is also some correlation with the galactic dust template from \cite{Jung:2018rgf}, but as the analysis was performed on the cleaned CMB map from which the dust has been removed, this has no impact on our final results.

\begin{table}
\begin{center}
\begin{tabular}{ |c|c|c|c| } 
 \hline
  & bouncing & bouncing & bouncing \\ 
  & ($q=-0.5$) & ($q=-0.7$) & ($q=-1.24$) \\ 
 \hline \hline
local & 0.018 & 0.013 & 0.006\\
 \hline
 equilateral & 0.011 & 0.006 & -0.002 \\
 \hline
 orthogonal & -0.046 & -0.039 & -0.028 \\
 \hline
 point sources & -$10^{-10}$ & -$10^{-10}$ & -$10^{-11}$ \\
 \hline
 CIB & -$10^{-7}$ & -$10^{-7}$ & -$10^{-8}$ \\
 \hline
 galactic dust & -0.13 & -0.11 & -0.066 \\
 \hline
 lensing & -0.002 & -0.002 & -0.001 \\
 \hline
 bouncing ($q=-0.5$) & & 0.98 &  0.82 \\
 \hline
 bouncing ($q=-0.7$) & & & 0.91 \\
 \hline
\end{tabular}
\caption{Correlation coefficients of the bouncing template (for the three different values of $q$) with the standard primordial and foreground bispectrum templates of the Planck analysis~\cite{Planck-nongaussian}, as well as with the galactic dust bispectrum template from~\cite{Jung:2018rgf}.}
\label{overlap}
\end{center}
\end{table}

Table~\ref{fNLresults} presents the final results for the analysis of the Planck 2018 SMICA CMB maps with the bouncing template. They have been computed using a full temperature plus E-mode polarization analysis. However, the addition of polarization does not help at all, one obtains exactly the same error bars using temperature only. We see that there is no detection of any of the three templates. Given the size of the error bars in this Table and the predicted values of $\fNL$ (20\%) given in Table~\ref{table:1}, we see that 
model~1 with $q=-0.5$ is ruled out at $5.4\sigma$,
the LQC model with $q=-0.7$ is ruled out at $6.4\sigma$, while model~3 with $q=-1.24$ is ruled out at $14\sigma$.

\begin{table}
\begin{center}
\begin{tabular}{ |c|c| } 
 \hline
  template & $\fNL$   \\ 
 \hline \hline
bouncing ($q=-0.5$) & ~~240 $\pm$ 470~~ \\
 \hline
 bouncing ($q=-0.7$) & ~~160 $\pm$ 260~~ \\
 \hline
 ~~bouncing ($q=-1.24$)~~ & 19 $\pm$ 34 \\
 \hline
\end{tabular}
\caption{$\fNL$ (with $1\sigma$ error bars) of the bouncing template (for the three different values of $q$) as determined from the 2018 Planck SMICA CMB temperature and polarization maps using the binned bispectrum estimator.} 
\label{fNLresults}
\end{center}
\end{table}
\vspace{0.2cm}

{\it Conclusion~~~ }
In this letter we have compared the non-Gaussianities of three bouncing models, which  mitigate the large-scale anomalies in the CMB data. Despite the fact that the bispectrum of these models decays exponentially below the pivot scale, for $k>k_b=0.002$~Mpc$^{-1}$, these models are excluded by the Planck data with high significance. This shows the sensitivity of the Planck data to scales beyond the pivot scale. This is especially evident when comparing models~2 and 3. While the LQC model has much larger $\fNL$ and therefore a larger bispectrum on all scales $k>k_b$, it is less significantly excluded, namely by $6.4\si$, than the third model with $q=-1.24$ which is excluded at $14\si$. The bispectrum of this model is smaller than the one from LQC for $k>k_b$, but is larger for $k<k_b/3.3$. These large scales are imprinted in the CMB since the CMB transfer function is by no means a Dirac delta and a given $\ell$ value obtains contributions from a rather broad band of wave numbers $k$. 

As lowering $\fNL$ in these models goes in pair with rendering $q$ even more negative, this implies that solving the large-scale anomaly puzzle with these models  is excluded by the Planck data.

If one reduces the probability  for the large-scale anomalies to appear from 20\% to 10 \% or even 5\%, this reduces the exclusion by the same factor as $\fNL$, see Table~\ref{table:1}, leading to only $3.5\si$ or $2.7\si$ exclusion for model~1 but still $9.3\si$ and $7.2\si$ for model~3. For the LQC model~2 the corresponding limits are $4.2\si$ and $3.3\si$, respectively.

It is very likely that our results actually go beyond the models studied here. If we want the large-scale anomalies to be less improbable by  skewed statistics, this introduces a bispectrum. Even if this bispectrum is significant only on very large scales, the Planck data are sufficiently precise to exclude it. It is of course possible that this might be evaded by some very exceptional, faster than exponential decay of the bispectrum; nevertheless, ours does appear to be a quite solid conclusion.

\begin{acknowledgments}\vspace{0.2cm} ~~ \\
We thank Nelson Pinto-Neto for helpful discussions. P.C.M.D. is supported by the grant No. UMO-2018/30/Q/ST9/00795 from the National Science Centre, Poland.
R.D. acknowledges support from the Swiss National Science Foundation, grant No. 200020\underline{~}182044. We gratefully acknowledge the IN2P3 Computer Centre (\url{https://cc.in2p3.fr}) for providing the computing resources and services needed for the analysis.

\end{acknowledgments}

\bibliography{main}

\begin{thebibliography}{33}%
\makeatletter
\providecommand \@ifxundefined [1]{%
 \@ifx{#1\undefined}
}%
\providecommand \@ifnum [1]{%
 \ifnum #1\expandafter \@firstoftwo
 \else \expandafter \@secondoftwo
 \fi
}%
\providecommand \@ifx [1]{%
 \ifx #1\expandafter \@firstoftwo
 \else \expandafter \@secondoftwo
 \fi
}%
\providecommand \natexlab [1]{#1}%
\providecommand \enquote  [1]{``#1''}%
\providecommand \bibnamefont  [1]{#1}%
\providecommand \bibfnamefont [1]{#1}%
\providecommand \citenamefont [1]{#1}%
\providecommand \href@noop [0]{\@secondoftwo}%
\providecommand \href [0]{\begingroup \@sanitize@url \@href}%
\providecommand \@href[1]{\@@startlink{#1}\@@href}%
\providecommand \@@href[1]{\endgroup#1\@@endlink}%
\providecommand \@sanitize@url [0]{\catcode `\\12\catcode `\$12\catcode
  `\&12\catcode `\#12\catcode `\^12\catcode `\_12\catcode `\%12\relax}%
\providecommand \@@startlink[1]{}%
\providecommand \@@endlink[0]{}%
\providecommand \url  [0]{\begingroup\@sanitize@url \@url }%
\providecommand \@url [1]{\endgroup\@href {#1}{\urlprefix }}%
\providecommand \urlprefix  [0]{URL }%
\providecommand \Eprint [0]{\href }%
\providecommand \doibase [0]{http://dx.doi.org/}%
\providecommand \selectlanguage [0]{\@gobble}%
\providecommand \bibinfo  [0]{\@secondoftwo}%
\providecommand \bibfield  [0]{\@secondoftwo}%
\providecommand \translation [1]{[#1]}%
\providecommand \BibitemOpen [0]{}%
\providecommand \bibitemStop [0]{}%
\providecommand \bibitemNoStop [0]{.\EOS\space}%
\providecommand \EOS [0]{\spacefactor3000\relax}%
\providecommand \BibitemShut  [1]{\csname bibitem#1\endcsname}%
\let\auto@bib@innerbib\@empty
\bibitem [{\citenamefont {Starobinsky}(1979)}]{inflation1}%
  \BibitemOpen
  \bibfield  {author} {\bibinfo {author} {\bibfnamefont {A.}~\bibnamefont
  {Starobinsky}},\ }\href@noop {} {\bibfield  {journal} {\bibinfo  {journal}
  {JETP Lett.}\ }\textbf {\bibinfo {volume} {30}},\ \bibinfo {pages} {682}
  (\bibinfo {year} {1979})}\BibitemShut {NoStop}%
\bibitem [{\citenamefont {Mukhanov}\ and\ \citenamefont
  {Chibisov}(1981)}]{inflation4}%
  \BibitemOpen
  \bibfield  {author} {\bibinfo {author} {\bibfnamefont {V.}~\bibnamefont
  {Mukhanov}}\ and\ \bibinfo {author} {\bibfnamefont {G.}~\bibnamefont
  {Chibisov}},\ }\href@noop {} {\bibfield  {journal} {\bibinfo  {journal} {JETP
  Lett.}\ }\textbf {\bibinfo {volume} {33}},\ \bibinfo {pages} {532} (\bibinfo
  {year} {1981})}\BibitemShut {NoStop}%
\bibitem [{\citenamefont {Mukhanov}\ and\ \citenamefont
  {Chibisov}(1982)}]{Mukhanov:1982nu}%
  \BibitemOpen
  \bibfield  {author} {\bibinfo {author} {\bibfnamefont {V.~F.}\ \bibnamefont
  {Mukhanov}}\ and\ \bibinfo {author} {\bibfnamefont {G.~V.}\ \bibnamefont
  {Chibisov}},\ }\href@noop {} {\bibfield  {journal} {\bibinfo  {journal} {Sov.
  Phys. JETP}\ }\textbf {\bibinfo {volume} {56}},\ \bibinfo {pages} {258}
  (\bibinfo {year} {1982})}\BibitemShut {NoStop}%
\bibitem [{\citenamefont {Guth}(1981)}]{inflation2}%
  \BibitemOpen
  \bibfield  {author} {\bibinfo {author} {\bibfnamefont {A.}~\bibnamefont
  {Guth}},\ }\href {\doibase 10.1103/PhysRevD.23.347} {\bibfield  {journal}
  {\bibinfo  {journal} {Phys. Rev. D}\ }\textbf {\bibinfo {volume} {23}},\
  \bibinfo {pages} {347} (\bibinfo {year} {1981})}\BibitemShut {NoStop}%
\bibitem [{\citenamefont {Linde}(1982)}]{inflation3}%
  \BibitemOpen
  \bibfield  {author} {\bibinfo {author} {\bibfnamefont {A.}~\bibnamefont
  {Linde}},\ }\href {\doibase 10.1016/0370-2693(82)91219-9} {\bibfield
  {journal} {\bibinfo  {journal} {Phys. Lett. B}\ }\textbf {\bibinfo {volume}
  {108}},\ \bibinfo {pages} {389} (\bibinfo {year} {1982})}\BibitemShut
  {NoStop}%
\bibitem [{\citenamefont {Tristram}\ \emph {et~al.}(2022)\citenamefont
  {Tristram} \emph {et~al.}}]{Tristram:2021tvh}%
  \BibitemOpen
  \bibfield  {author} {\bibinfo {author} {\bibfnamefont {M.}~\bibnamefont
  {Tristram}} \emph {et~al.},\ }\href {\doibase 10.1103/PhysRevD.105.083524}
  {\bibfield  {journal} {\bibinfo  {journal} {Phys. Rev. D}\ }\textbf {\bibinfo
  {volume} {105}},\ \bibinfo {pages} {083524} (\bibinfo {year} {2022})},\
  \Eprint {http://arxiv.org/abs/2112.07961} {arXiv:2112.07961 [astro-ph.CO]}
  \BibitemShut {NoStop}%
\bibitem [{\citenamefont {Bezrukov}\ and\ \citenamefont
  {Shaposhnikov}(2008)}]{Bezrukov:2007ep}%
  \BibitemOpen
  \bibfield  {author} {\bibinfo {author} {\bibfnamefont {F.~L.}\ \bibnamefont
  {Bezrukov}}\ and\ \bibinfo {author} {\bibfnamefont {M.}~\bibnamefont
  {Shaposhnikov}},\ }\href {\doibase 10.1016/j.physletb.2007.11.072} {\bibfield
   {journal} {\bibinfo  {journal} {Phys. Lett. B}\ }\textbf {\bibinfo {volume}
  {659}},\ \bibinfo {pages} {703} (\bibinfo {year} {2008})},\ \Eprint
  {http://arxiv.org/abs/0710.3755} {arXiv:0710.3755 [hep-th]} \BibitemShut
  {NoStop}%
\bibitem [{\citenamefont {Tolman}(1931)}]{cbounce1}%
  \BibitemOpen
  \bibfield  {author} {\bibinfo {author} {\bibfnamefont {R.}~\bibnamefont
  {Tolman}},\ }\href {\doibase 10.1103/PhysRev.38.1758} {\bibfield  {journal}
  {\bibinfo  {journal} {Phys. Rev.}\ }\textbf {\bibinfo {volume} {38}},\
  \bibinfo {pages} {1758} (\bibinfo {year} {1931})}\BibitemShut {NoStop}%
\bibitem [{\citenamefont {Murphy}(1973)}]{cbounce2}%
  \BibitemOpen
  \bibfield  {author} {\bibinfo {author} {\bibfnamefont {G.}~\bibnamefont
  {Murphy}},\ }\href {\doibase 10.1103/PhysRevD.8.4231} {\bibfield  {journal}
  {\bibinfo  {journal} {Phys. Rev.}\ }\textbf {\bibinfo {volume} {8}},\
  \bibinfo {pages} {4231} (\bibinfo {year} {1973})}\BibitemShut {NoStop}%
\bibitem [{\citenamefont {Easson}\ \emph {et~al.}(2011)\citenamefont {Easson},
  \citenamefont {Sawicki},\ and\ \citenamefont {Vikman}}]{Easson:2011zy}%
  \BibitemOpen
  \bibfield  {author} {\bibinfo {author} {\bibfnamefont {D.~A.}\ \bibnamefont
  {Easson}}, \bibinfo {author} {\bibfnamefont {I.}~\bibnamefont {Sawicki}}, \
  and\ \bibinfo {author} {\bibfnamefont {A.}~\bibnamefont {Vikman}},\ }\href
  {\doibase 10.1088/1475-7516/2011/11/021} {\bibfield  {journal} {\bibinfo
  {journal} {JCAP}\ }\textbf {\bibinfo {volume} {11}},\ \bibinfo {pages} {021}
  (\bibinfo {year} {2011})},\ \Eprint {http://arxiv.org/abs/1109.1047}
  {arXiv:1109.1047 [hep-th]} \BibitemShut {NoStop}%
\bibitem [{\citenamefont {Fabris}\ \emph {et~al.}(2012)\citenamefont {Fabris},
  \citenamefont {Perez}, \citenamefont {Bergliaffa},\ and\ \citenamefont
  {Pinto-Neto}}]{cbounce6}%
  \BibitemOpen
  \bibfield  {author} {\bibinfo {author} {\bibfnamefont {J.}~\bibnamefont
  {Fabris}}, \bibinfo {author} {\bibfnamefont {R.}~\bibnamefont {Perez}},
  \bibinfo {author} {\bibfnamefont {S.}~\bibnamefont {Bergliaffa}}, \ and\
  \bibinfo {author} {\bibfnamefont {N.}~\bibnamefont {Pinto-Neto}},\ }\href
  {\doibase 10.1103/PhysRevD.86.103525} {\bibfield  {journal} {\bibinfo
  {journal} {Phys. Rev. D}\ }\textbf {\bibinfo {volume} {86}},\ \bibinfo
  {pages} {103525} (\bibinfo {year} {2012})},\ \Eprint
  {http://arxiv.org/abs/1205.3458} {arXiv:1205.3458 [astro-ph.CO]} \BibitemShut
  {NoStop}%
\bibitem [{\citenamefont {Ijjas}\ and\ \citenamefont
  {Steinhardt}(2016{\natexlab{a}})}]{cbounce8}%
  \BibitemOpen
  \bibfield  {author} {\bibinfo {author} {\bibfnamefont {A.}~\bibnamefont
  {Ijjas}}\ and\ \bibinfo {author} {\bibfnamefont {P.}~\bibnamefont
  {Steinhardt}},\ }\href {\doibase 10.1103/PhysRevLett.117.121304} {\bibfield
  {journal} {\bibinfo  {journal} {Phys. Rev. Lett.}\ }\textbf {\bibinfo
  {volume} {117}},\ \bibinfo {pages} {121304} (\bibinfo {year}
  {2016}{\natexlab{a}})},\ \Eprint {http://arxiv.org/abs/1606.08880}
  {arXiv:1606.08880 [gr-qc]} \BibitemShut {NoStop}%
\bibitem [{\citenamefont {Dobre}\ \emph {et~al.}(2018)\citenamefont {Dobre},
  \citenamefont {Frolov}, \citenamefont {G\'alvez~Ghersi}, \citenamefont
  {Ramazanov},\ and\ \citenamefont {Vikman}}]{Dobre:2017pnt}%
  \BibitemOpen
  \bibfield  {author} {\bibinfo {author} {\bibfnamefont {D.~A.}\ \bibnamefont
  {Dobre}}, \bibinfo {author} {\bibfnamefont {A.~V.}\ \bibnamefont {Frolov}},
  \bibinfo {author} {\bibfnamefont {J.~T.}\ \bibnamefont {G\'alvez~Ghersi}},
  \bibinfo {author} {\bibfnamefont {S.}~\bibnamefont {Ramazanov}}, \ and\
  \bibinfo {author} {\bibfnamefont {A.}~\bibnamefont {Vikman}},\ }\href
  {\doibase 10.1088/1475-7516/2018/03/020} {\bibfield  {journal} {\bibinfo
  {journal} {JCAP}\ }\textbf {\bibinfo {volume} {03}},\ \bibinfo {pages} {020}
  (\bibinfo {year} {2018})},\ \Eprint {http://arxiv.org/abs/1712.10272}
  {arXiv:1712.10272 [gr-qc]} \BibitemShut {NoStop}%
\bibitem [{\citenamefont {Bojowald}(2001)}]{qbounce5}%
  \BibitemOpen
  \bibfield  {author} {\bibinfo {author} {\bibfnamefont {M.}~\bibnamefont
  {Bojowald}},\ }\href {\doibase 10.1103/PhysRevLett.86.5227} {\bibfield
  {journal} {\bibinfo  {journal} {Phys. Rev. Lett.}\ }\textbf {\bibinfo
  {volume} {86}},\ \bibinfo {pages} {5227} (\bibinfo {year} {2001})},\ \Eprint
  {http://arxiv.org/abs/:gr-qc/0102069} {arXiv::gr-qc/0102069} \BibitemShut
  {NoStop}%
\bibitem [{\citenamefont {Ashtekar}\ and\ \citenamefont
  {Singh}(2011)}]{qbounce10}%
  \BibitemOpen
  \bibfield  {author} {\bibinfo {author} {\bibfnamefont {A.}~\bibnamefont
  {Ashtekar}}\ and\ \bibinfo {author} {\bibfnamefont {P.}~\bibnamefont
  {Singh}},\ }\href {\doibase 10.1088/0264-9381/28/21/213001} {\bibfield
  {journal} {\bibinfo  {journal} {Class. Quant. Grav.}\ }\textbf {\bibinfo
  {volume} {28}},\ \bibinfo {pages} {213001} (\bibinfo {year} {2011})},\
  \Eprint {http://arxiv.org/abs/1108.0893} {arXiv:1108.0893 [gr-qc]}
  \BibitemShut {NoStop}%
\bibitem [{\citenamefont {Ijjas}\ and\ \citenamefont
  {Steinhardt}(2016{\natexlab{b}})}]{ob0}%
  \BibitemOpen
  \bibfield  {author} {\bibinfo {author} {\bibfnamefont {A.}~\bibnamefont
  {Ijjas}}\ and\ \bibinfo {author} {\bibfnamefont {P.}~\bibnamefont
  {Steinhardt}},\ }\href {\doibase 10.1088/0264-9381/33/4/044001/pdf}
  {\bibfield  {journal} {\bibinfo  {journal} {Class. Quantum Grav.}\ }\textbf
  {\bibinfo {volume} {33}},\ \bibinfo {pages} {044001} (\bibinfo {year}
  {2016}{\natexlab{b}})},\ \Eprint {http://arxiv.org/abs/1512.09010}
  {arXiv:1512.09010 [astro-ph.CO]} \BibitemShut {NoStop}%
\bibitem [{\citenamefont {Bacalhau}\ \emph {et~al.}(2018)\citenamefont
  {Bacalhau}, \citenamefont {Pinto-Neto},\ and\ \citenamefont {Vitenti}}]{ob2}%
  \BibitemOpen
  \bibfield  {author} {\bibinfo {author} {\bibfnamefont {A.}~\bibnamefont
  {Bacalhau}}, \bibinfo {author} {\bibfnamefont {N.}~\bibnamefont
  {Pinto-Neto}}, \ and\ \bibinfo {author} {\bibfnamefont {S.}~\bibnamefont
  {Vitenti}},\ }\href {\doibase 10.1103/PhysRevD.97.083517} {\bibfield
  {journal} {\bibinfo  {journal} {Phys. Rev. D}\ }\textbf {\bibinfo {volume}
  {97}},\ \bibinfo {pages} {083517} (\bibinfo {year} {2018})},\ \Eprint
  {http://arxiv.org/abs/1706.08830} {arXiv:1706.08830 [gr-qc]} \BibitemShut
  {NoStop}%
\bibitem [{\citenamefont {Agullo}\ \emph {et~al.}(2018)\citenamefont {Agullo},
  \citenamefont {Bolliet},\ and\ \citenamefont
  {Sreenath}}]{PhysRevD.97.066021}%
  \BibitemOpen
  \bibfield  {author} {\bibinfo {author} {\bibfnamefont {I.}~\bibnamefont
  {Agullo}}, \bibinfo {author} {\bibfnamefont {B.}~\bibnamefont {Bolliet}}, \
  and\ \bibinfo {author} {\bibfnamefont {V.}~\bibnamefont {Sreenath}},\ }\href
  {\doibase 10.1103/PhysRevD.97.066021} {\bibfield  {journal} {\bibinfo
  {journal} {Phys. Rev. D}\ }\textbf {\bibinfo {volume} {97}},\ \bibinfo
  {pages} {066021} (\bibinfo {year} {2018})},\ \Eprint
  {http://arxiv.org/abs/1712.08148} {arXiv:1712.08148 [gr-qc]} \BibitemShut
  {NoStop}%
\bibitem [{\citenamefont {Kim}\ and\ \citenamefont
  {Naselsky}(2011)}]{Kim:2010st}%
  \BibitemOpen
  \bibfield  {author} {\bibinfo {author} {\bibfnamefont {J.}~\bibnamefont
  {Kim}}\ and\ \bibinfo {author} {\bibfnamefont {P.}~\bibnamefont {Naselsky}},\
  }\href {\doibase 10.1088/0004-637X/739/2/79} {\bibfield  {journal} {\bibinfo
  {journal} {Astrophys. J.}\ }\textbf {\bibinfo {volume} {739}},\ \bibinfo
  {pages} {79} (\bibinfo {year} {2011})},\ \Eprint
  {http://arxiv.org/abs/1011.0377} {arXiv:1011.0377 [astro-ph.CO]} \BibitemShut
  {NoStop}%
\bibitem [{\citenamefont {Schwarz}\ \emph {et~al.}(2016)\citenamefont
  {Schwarz}, \citenamefont {Copi}, \citenamefont {Huterer},\ and\ \citenamefont
  {Starkman}}]{anomalies}%
  \BibitemOpen
  \bibfield  {author} {\bibinfo {author} {\bibfnamefont {D.}~\bibnamefont
  {Schwarz}}, \bibinfo {author} {\bibfnamefont {C.~J.}\ \bibnamefont {Copi}},
  \bibinfo {author} {\bibfnamefont {D.}~\bibnamefont {Huterer}}, \ and\
  \bibinfo {author} {\bibfnamefont {G.}~\bibnamefont {Starkman}},\ }\href
  {\doibase 10.1088/0264-9381/33/18/184001/meta} {\bibfield  {journal}
  {\bibinfo  {journal} {Class. Quant. Grav.}\ }\textbf {\bibinfo {volume}
  {33}},\ \bibinfo {pages} {184001} (\bibinfo {year} {2016})},\ \Eprint
  {http://arxiv.org/abs/1510.07929} {arXiv:1510.07929 [astro-ph.CO]}
  \BibitemShut {NoStop}%
\bibitem [{\citenamefont {Akrami}\ \emph
  {et~al.}(2020{\natexlab{a}})\citenamefont {Akrami} \emph
  {et~al.}}]{Planck-nongaussian}%
  \BibitemOpen
  \bibfield  {author} {\bibinfo {author} {\bibfnamefont {Y.}~\bibnamefont
  {Akrami}} \emph {et~al.} (\bibinfo {collaboration} {Planck}),\ }\href
  {\doibase 10.1051/0004-6361/201935891} {\bibfield  {journal} {\bibinfo
  {journal} {Astron. Astrophys.}\ }\textbf {\bibinfo {volume} {641}},\ \bibinfo
  {pages} {A9} (\bibinfo {year} {2020}{\natexlab{a}})},\ \Eprint
  {http://arxiv.org/abs/1905.05697} {arXiv:1905.05697 [astro-ph.CO]}
  \BibitemShut {NoStop}%
\bibitem [{\citenamefont {Agullo}\ \emph {et~al.}(2021)\citenamefont {Agullo},
  \citenamefont {Kranas},\ and\ \citenamefont {Sreenath}}]{Agullo:2020cvg}%
  \BibitemOpen
  \bibfield  {author} {\bibinfo {author} {\bibfnamefont {I.}~\bibnamefont
  {Agullo}}, \bibinfo {author} {\bibfnamefont {D.}~\bibnamefont {Kranas}}, \
  and\ \bibinfo {author} {\bibfnamefont {V.}~\bibnamefont {Sreenath}},\ }\href
  {\doibase 10.1088/1361-6382/abc521} {\bibfield  {journal} {\bibinfo
  {journal} {Class. Quant. Grav.}\ }\textbf {\bibinfo {volume} {38}},\ \bibinfo
  {pages} {065010} (\bibinfo {year} {2021})},\ \Eprint
  {http://arxiv.org/abs/2006.09605} {arXiv:2006.09605 [astro-ph.CO]}
  \BibitemShut {NoStop}%
\bibitem [{\citenamefont {Delgado}\ \emph {et~al.}(2021)\citenamefont
  {Delgado}, \citenamefont {Durrer},\ and\ \citenamefont
  {Pinto-Neto}}]{Delgado:2021mxu}%
  \BibitemOpen
  \bibfield  {author} {\bibinfo {author} {\bibfnamefont {P.~C.~M.}\
  \bibnamefont {Delgado}}, \bibinfo {author} {\bibfnamefont {R.}~\bibnamefont
  {Durrer}}, \ and\ \bibinfo {author} {\bibfnamefont {N.}~\bibnamefont
  {Pinto-Neto}},\ }\href {\doibase 10.1088/1475-7516/2021/11/024} {\bibfield
  {journal} {\bibinfo  {journal} {JCAP}\ }\textbf {\bibinfo {volume} {11}},\
  \bibinfo {pages} {024} (\bibinfo {year} {2021})},\ \Eprint
  {http://arxiv.org/abs/2108.06175} {arXiv:2108.06175 [astro-ph.CO]}
  \BibitemShut {NoStop}%
\bibitem [{\citenamefont {Bucher}\ \emph {et~al.}(2010)\citenamefont {Bucher},
  \citenamefont {Van~Tent},\ and\ \citenamefont {Carvalho}}]{Bucher:2009nm}%
  \BibitemOpen
  \bibfield  {author} {\bibinfo {author} {\bibfnamefont {M.}~\bibnamefont
  {Bucher}}, \bibinfo {author} {\bibfnamefont {B.}~\bibnamefont {Van~Tent}}, \
  and\ \bibinfo {author} {\bibfnamefont {C.~S.}\ \bibnamefont {Carvalho}},\
  }\href {\doibase 10.1111/j.1365-2966.2010.17089.x} {\bibfield  {journal}
  {\bibinfo  {journal} {Mon. Not. Roy. Astron. Soc.}\ }\textbf {\bibinfo
  {volume} {407}},\ \bibinfo {pages} {2193} (\bibinfo {year} {2010})},\ \Eprint
  {http://arxiv.org/abs/0911.1642} {arXiv:0911.1642 [astro-ph.CO]} \BibitemShut
  {NoStop}%
\bibitem [{\citenamefont {Bucher}\ \emph {et~al.}(2016)\citenamefont {Bucher},
  \citenamefont {Racine},\ and\ \citenamefont {van Tent}}]{Bucher:2015ura}%
  \BibitemOpen
  \bibfield  {author} {\bibinfo {author} {\bibfnamefont {M.}~\bibnamefont
  {Bucher}}, \bibinfo {author} {\bibfnamefont {B.}~\bibnamefont {Racine}}, \
  and\ \bibinfo {author} {\bibfnamefont {B.}~\bibnamefont {van Tent}},\ }\href
  {\doibase 10.1088/1475-7516/2016/05/055} {\bibfield  {journal} {\bibinfo
  {journal} {JCAP}\ }\textbf {\bibinfo {volume} {05}},\ \bibinfo {pages} {055}
  (\bibinfo {year} {2016})},\ \Eprint {http://arxiv.org/abs/1509.08107}
  {arXiv:1509.08107 [astro-ph.CO]} \BibitemShut {NoStop}%
\bibitem [{\citenamefont {Komatsu}\ and\ \citenamefont
  {Spergel}(2001)}]{Komatsu:2001rj}%
  \BibitemOpen
  \bibfield  {author} {\bibinfo {author} {\bibfnamefont {E.}~\bibnamefont
  {Komatsu}}\ and\ \bibinfo {author} {\bibfnamefont {D.~N.}\ \bibnamefont
  {Spergel}},\ }\href {\doibase 10.1103/PhysRevD.63.063002} {\bibfield
  {journal} {\bibinfo  {journal} {Phys. Rev. D}\ }\textbf {\bibinfo {volume}
  {63}},\ \bibinfo {pages} {063002} (\bibinfo {year} {2001})},\ \Eprint
  {http://arxiv.org/abs/astro-ph/0005036} {arXiv:astro-ph/0005036} \BibitemShut
  {NoStop}%
\bibitem [{\citenamefont {Aghanim}\ \emph {et~al.}(2020)\citenamefont {Aghanim}
  \emph {et~al.}}]{Planck:2018vyg}%
  \BibitemOpen
  \bibfield  {author} {\bibinfo {author} {\bibfnamefont {N.}~\bibnamefont
  {Aghanim}} \emph {et~al.} (\bibinfo {collaboration} {Planck}),\ }\href
  {\doibase 10.1051/0004-6361/201833910} {\bibfield  {journal} {\bibinfo
  {journal} {Astron. Astrophys.}\ }\textbf {\bibinfo {volume} {641}},\ \bibinfo
  {pages} {A6} (\bibinfo {year} {2020})},\ \Eprint
  {http://arxiv.org/abs/1807.06209} {arXiv:1807.06209 [astro-ph.CO]}
  \BibitemShut {NoStop}%
\bibitem [{\citenamefont {Durrer}(2020)}]{Durrer:2020fza}%
  \BibitemOpen
  \bibfield  {author} {\bibinfo {author} {\bibfnamefont {R.}~\bibnamefont
  {Durrer}},\ }\href {\doibase 10.1017/9781316471524} {\emph {\bibinfo {title}
  {{The Cosmic Microwave Background}}}}\ (\bibinfo  {publisher} {Cambridge
  University Press},\ \bibinfo {year} {2020})\BibitemShut {NoStop}%
\bibitem [{\citenamefont {Hu}\ and\ \citenamefont {Okamoto}(2004)}]{Hu:2003vp}%
  \BibitemOpen
  \bibfield  {author} {\bibinfo {author} {\bibfnamefont {W.}~\bibnamefont
  {Hu}}\ and\ \bibinfo {author} {\bibfnamefont {T.}~\bibnamefont {Okamoto}},\
  }\href {\doibase 10.1103/PhysRevD.69.043004} {\bibfield  {journal} {\bibinfo
  {journal} {Phys. Rev. D}\ }\textbf {\bibinfo {volume} {69}},\ \bibinfo
  {pages} {043004} (\bibinfo {year} {2004})},\ \Eprint
  {http://arxiv.org/abs/astro-ph/0308049} {arXiv:astro-ph/0308049} \BibitemShut
  {NoStop}%
\bibitem [{\citenamefont {Ade}\ \emph {et~al.}(2014)\citenamefont {Ade} \emph
  {et~al.}}]{Planck:2013wtn}%
  \BibitemOpen
  \bibfield  {author} {\bibinfo {author} {\bibfnamefont {P.~A.~R.}\
  \bibnamefont {Ade}} \emph {et~al.} (\bibinfo {collaboration} {Planck}),\
  }\href {\doibase 10.1051/0004-6361/201321554} {\bibfield  {journal} {\bibinfo
   {journal} {Astron. Astrophys.}\ }\textbf {\bibinfo {volume} {571}},\
  \bibinfo {pages} {A24} (\bibinfo {year} {2014})},\ \Eprint
  {http://arxiv.org/abs/1303.5084} {arXiv:1303.5084 [astro-ph.CO]} \BibitemShut
  {NoStop}%
\bibitem [{\citenamefont {Ade}\ \emph {et~al.}(2016)\citenamefont {Ade} \emph
  {et~al.}}]{Planck:2016aaa}%
  \BibitemOpen
  \bibfield  {author} {\bibinfo {author} {\bibfnamefont {P.~A.~R.}\
  \bibnamefont {Ade}} \emph {et~al.} (\bibinfo {collaboration} {Planck}),\
  }\href {\doibase 10.1051/0004-6361/201525836} {\bibfield  {journal} {\bibinfo
   {journal} {Astron. Astrophys.}\ }\textbf {\bibinfo {volume} {594}},\
  \bibinfo {pages} {A17} (\bibinfo {year} {2016})},\ \Eprint
  {http://arxiv.org/abs/1502.01592} {arXiv:1502.01592 [astro-ph.CO]}
  \BibitemShut {NoStop}%
\bibitem [{\citenamefont {Akrami}\ \emph
  {et~al.}(2020{\natexlab{b}})\citenamefont {Akrami} \emph
  {et~al.}}]{Planck:2018yye}%
  \BibitemOpen
  \bibfield  {author} {\bibinfo {author} {\bibfnamefont {Y.}~\bibnamefont
  {Akrami}} \emph {et~al.} (\bibinfo {collaboration} {Planck}),\ }\href
  {\doibase 10.1051/0004-6361/201833881} {\bibfield  {journal} {\bibinfo
  {journal} {Astron. Astrophys.}\ }\textbf {\bibinfo {volume} {641}},\ \bibinfo
  {pages} {A4} (\bibinfo {year} {2020}{\natexlab{b}})},\ \Eprint
  {http://arxiv.org/abs/1807.06208} {arXiv:1807.06208 [astro-ph.CO]}
  \BibitemShut {NoStop}%
\bibitem [{\citenamefont {Jung}\ \emph {et~al.}(2018)\citenamefont {Jung},
  \citenamefont {Racine},\ and\ \citenamefont {van Tent}}]{Jung:2018rgf}%
  \BibitemOpen
  \bibfield  {author} {\bibinfo {author} {\bibfnamefont {G.}~\bibnamefont
  {Jung}}, \bibinfo {author} {\bibfnamefont {B.}~\bibnamefont {Racine}}, \ and\
  \bibinfo {author} {\bibfnamefont {B.}~\bibnamefont {van Tent}},\ }\href
  {\doibase 10.1088/1475-7516/2018/11/047} {\bibfield  {journal} {\bibinfo
  {journal} {JCAP}\ }\textbf {\bibinfo {volume} {11}},\ \bibinfo {pages} {047}
  (\bibinfo {year} {2018})},\ \Eprint {http://arxiv.org/abs/1810.01727}
  {arXiv:1810.01727 [astro-ph.CO]} \BibitemShut {NoStop}%
\end{thebibliography}%

\end{document}